\documentclass[aps,pre,preprint,groupedaddress,floatfix,longbibliography]{revtex4-1}

\usepackage{amsmath}
\usepackage{graphicx}
\usepackage{color}

\begin{document}


\title{Waveguide metacouplers for in-plane polarimetry}


\author{Anders Pors}
\email[]{alp@iti.sdu.dk}
\affiliation{SDU Nano Optics, University of Southern Denmark, Campusvej 55, DK-5230 Odense M, Denmark}

\author{Sergey I. Bozhevolnyi}
\affiliation{SDU Nano Optics, University of Southern Denmark, Campusvej 55, DK-5230 Odense M, Denmark}


\date{\today}

\begin{abstract}
The state of polarization (SOP) is an inherent property of the vectorial nature of light and a crucial parameter in a wide range of remote sensing applications. Nevertheless, the SOP is rather cumbersome to probe experimentally, as conventional detectors only respond to the intensity of the light, hence loosing the phase information between orthogonal vector components. In this work, we propose a new type of polarimeter that is compact and well-suited for in-plane optical circuitry, while allowing for immediate determination of the SOP through simultaneous retrieval of the associated Stokes parameters. The polarimeter is based on plasmonic phase-gradient birefringent metasurfaces that facilitate normal incident light to launch in-plane photonic waveguide modes propagating in six predefined directions with the coupling efficiencies providing a direct measure of the incident SOP. The functionality and accuracy of the polarimeter, which essentially is an all-polarization sensitive waveguide metacoupler, is confirmed through full-wave simulations at the operation wavelength of $1.55$\,$\mu$m.
\end{abstract}


\maketitle

\section{Introduction}
Despite the fact that the state of polarization (SOP) is the key characteristic of the vectorial nature of electromagnetic waves, it is an inherently difficult parameter to experimentally probe owing to the loss of information on the relative phase between orthogonal vector components in conventional (intensity) detection schemes. Nevertheless, the SOP (or the change in SOP) is a parameter often sought measured since it may carry crucial information about the composition and structure of the medium that the wave has been interacting with. As prominent applications, we mention remote sensing within astronomy \cite{Tinbergen_1996}, biology \cite{Collings_1997}, and camouflage technology \cite{Tyo_2006}, but also more nascent applications, such as for the fundamental understanding of processes in laser fusion or within the field of quantum communication, advantageously exploit the knowledge of the SOP \cite{Lepetit_2015}. Overall, it transpires that polarimetry is of utmost importance in both fundamental and applied science. 

The SOP evaluation is typically based on determination of the so-called Stokes parameters, which are constructed from six intensity measurements with properly arranged polarizers placed in front of the detector, hereby allowing one to uniquely retrieve the SOP \cite{bohren}. We note that the series of measurements can be automated (as in commercial polarimeters), though at the expense of a non-negligible acquisition time that may induce errors or limit the possibility to measure transient events. Alternatively, the Stokes parameters can be measured simultaneously by splitting the beam into multiple optical paths and utilizing several polarizers and detectors \cite{Compain_1998}. The downside of this approach, however, amounts in a bulky, complex, and expense optical system. Also, we note that additional realizations of polarimeters do exist, like using advanced micro-polarizers in front of an imaging detector, but those approaches are typically complex and expensive \cite{Tyo_2006}. Overall, it seems that none of the conventional approaches is ideal with respect to simultaneous determination of the Stokes parameters, compact and inexpensive design, and ease of usage (e.g., no tedious alignment etc.). 

With the above outline of the current status of polarimeters, it is natural to discuss the recent advances within nanophotonics, particularly the new degrees of freedom in controlling light using metasurfaces \cite{Yu_2014,Estakhri_2016}. Here, early approaches in determining the SOP utilized a combination of a metasurface together with conventional optical elements \cite{Gori_99,Bomzon_01} (like polarizers and waveplates), or the effect of polarization-dependent transmission of light through six carefully designed nano-apertures in metal films \cite{Afshinmanesh_2012}. Likewise, different types of metasurfaces have been proposed for determination of certain aspects of the SOP, like the degree of linear \cite{Mandel_2013} or circular \cite{Wen_2015,Shaltout_2015} polarization. Recently, however, metasurface-only polarimeters that uniquely identify the SOP have been proposed and verified. For example, a nanometer-thin metadevice consisting of an array of meticulously designed (rotated and aligned) metallic nanoantennas will feature an in-plane scattering pattern that is unique for all SOPs \cite{Mueller_2016}. In a different study, we have proposed a reflective metagrating that redirects light into six diffraction orders, with the pair-wise contrast in diffraction intensities immediately revealing the Stokes parameters of the incident SOP \cite{Pors_2015}. Our polarimeter is based on the optical analog of the reflectarray concept \cite{Pozar_1997}, hence consisting of an optically thick metal film overlaid by a nanometer thin dielectric spacer and an array of carefully designed metallic nanobricks. These metasurfaces, also known as gap surface plasmon-based (GSP-based) metasurfaces, have the attractive property of enabling simultaneous control of either the amplitude and phase of the reflected light for one polarization or independently engineering the reflection phases for two orthogonal polarizations \cite{Pors_2013}. These new possibilities for light control have been exploited in metasurfaces performing analog computations on the incident light \cite{Pors_2015_2}, dual-image holograms \cite{Chen_2014}, and polarization-controlled unidirectional excitation of surface plasmon polaritons \cite{Pors_2014}. Particularly, the last application has inspired us to suggest a new type of compact metasurface-based polarimeter that couples incident light into in-plane waveguide modes, with the relative efficiency of excitation between predefined propagation directions being directly related to the SOP.  We incorporate in our design three GSP-based metasurfaces that unidirectionally excite the waveguide modes propagating in six different directions for the three different polarization bases that are dictated by the definition of Stokes parameters. As a way of example, we design the all-polarization sensitive waveguide metacoupler at a wavelength of $1.55$\,$\mu$m and perform full-wave numerical simulations of a realistic ($\sim 100$\,$\mu$m$^2$ footprint) device that reveals the possibility to accurately retrieve the Stokes parameters in one shot.

\section{Stokes parameters}
Before we begin discussing the realization of the in-plane polarimeter, it is appropriate to quickly review the connection between the polarization of a plane wave, described by the conventional Jones vector, and the Stokes parameters that are typically measured in experiments. For a $z$-propagating monochromatic plane wave, the Jones vector can be written as
\begin{equation}
\mathbf{E}_0=
\begin{pmatrix}
A_x \\
A_y e^{i\delta}	
\end{pmatrix},
\label{eq:E0}
\end{equation} 
where ($A_x,A_y$) are real-valued amplitude coefficients and $\delta$ describes the phase difference between those two components. Despite the simplicity in describing the amplitude and SOP mathematically, the latter parameter is in contrast inherently difficult to probe experimentally, which owes to the fact that conventional detectors respond to the intensity of the impinging wave (i.e., $I\propto A_x^2+A_y^2$), hence loosing information of the crucial phase-relation between the two orthogonal components. In order to remedy this shortcoming in experiments, the four Stokes parameters are introduced, which are based on six intensity measurements and fully describe both the amplitude and SOP of the plane wave. The Stokes parameters can be written as
\begin{align}
& s_0 =A_x^2+A_y^2, \\
\label{eq:s1}
& s_1 =A_x^2-A_y^2, \\
& s_2 =2A_xA_y\cos\delta=A_a^2-A_b^2, \\
& s_3 =2A_xA_y\sin\delta=A_r^2-A_l^2,
\label{eq:s3}
\end{align} 
where it is readily seen that $s_0$ simply describes the intensity of the beam, thus retaining the information of the SOP in $s_1-s_3$. Moreover, $s_1-s_3$ can be found by measuring the intensity of the two orthogonal components of the light in the three bases ($\hat{\mathbf{x}},\hat{\mathbf{y}}$), ($\hat{\mathbf{a}},\hat{\mathbf{b}}$)=$\tfrac{1}{\sqrt{2}}(\hat{\mathbf{x}}+\hat{\mathbf{y}},-\hat{\mathbf{x}}+\hat{\mathbf{y}})$, and ($\hat{\mathbf{r}},\hat{\mathbf{l}}$)=$\tfrac{1}{\sqrt{2}}(\hat{\mathbf{x}}+i\hat{\mathbf{y}},\hat{\mathbf{x}}-i\hat{\mathbf{y}})$, where the latter two bases correspond to rotation of the Cartesian coordinate system ($\hat{\mathbf{x}},\hat{\mathbf{y}}$) by $45^\circ$ with respect to the $x$-axis and the basis for circularly polarized light, respectively. It should be noted that in describing the SOP of a plane wave, $s_1-s_3$ are conventionally normalized by $s_0$ so that all possible values lie within $\pm 1$. Additionally, it is seen that $(s_1^2+s_2^2+s_3^2)/s_0^2=1$, which signifies that all polarization states in the three-dimensional space ($s_1,s_2,s_3$) represent a unit sphere, also known as the Poincar{\'e} sphere.

Having outlined the connection between SOP and the Stokes parameters, it is clear that our waveguide metacoupler must respond uniquely to \emph{all} possible SOPs, with preferably the most pronounced differences occurring for the six extreme polarizations $|x\rangle$, $|y\rangle$, $|a\rangle$, $|b\rangle$, $|r\rangle$, and $|l\rangle$, so that all linear polarizations thereof can be probably resolved. In order to achieve this property, we base our design on \emph{birefringent} GSP-based metasurfaces that can be used for unidirectional and polarization-controlled interfacing of freely propagating waves and waveguide modes \cite{Pors_2014}. The in-plane momentum matching to the waveguide mode is achieved through grating coupling, with an additional linear phase gradient along the metasurface ensuring unidirectional excitation. Moreover, and in line with the previous work \cite{Pors_2015}, the metacoupler will consist of three metasurfaces that launch the waveguide modes in different directions for the orthogonal sets of polarizations ($|x\rangle$,$|y\rangle$), ($|a\rangle$,$|b\rangle$), and ($|r\rangle$,$|l\rangle$), respectively. In this way, the contrast between the power carried by the waveguide mode in the two propagation directions of each metasurface will mimic the respective dependence of $s_1-s_3$ on the SOP. 

\section{Design of waveguide metacouplers} 
\begin{figure}[tb]
	\centering
		\includegraphics[width=8.5cm]{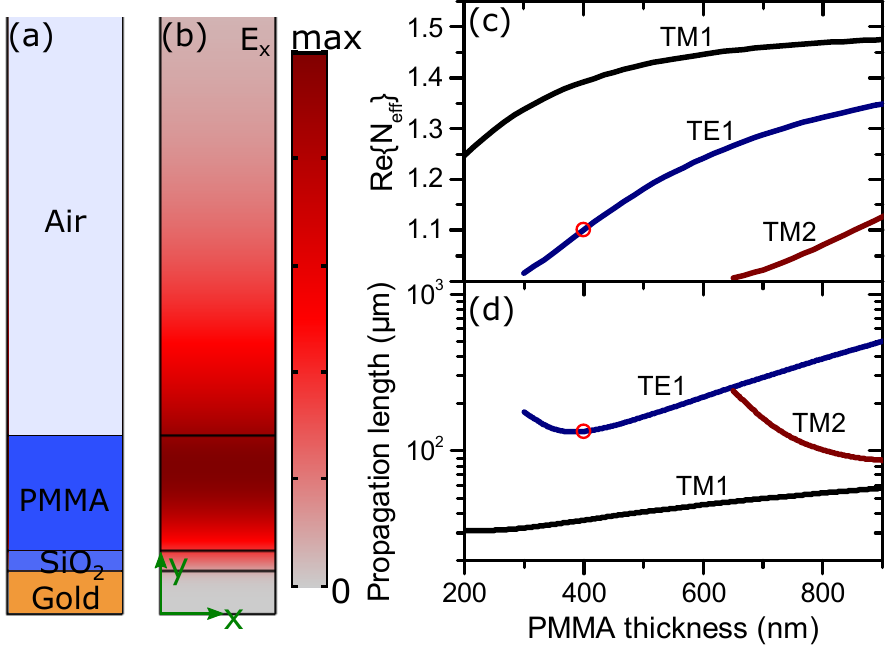}
	\caption{Waveguide configuration. (a) Sketch of waveguide configuration that is assumed spatial invariant along the $x$- and $z$-direction. (b) The electric field of TE1 mode for propagation along $z$-direction and PMMA thickness of 400\,nm. (c) The real part of the effective index $N_\mathrm{eff}=\beta/k_0$, where $\beta$ is the propagation constant of the mode and $k_0$ is the vacuum wave number, and (d) the propagation length for modes sustained by the configuration in a) as a function of PMMA thickness. The SiO$_2$-thickness is fixed at 70\,nm and the wavelength is $\lambda=1550$\,nm.}
	\label{fig:WaveguideMode}
\end{figure}

In the design of any waveguide coupler, the first step is to specify the properties of the mode to be launched by the coupler. The waveguide configuration considered here consists of an optically thick gold film overlaid by a 70\,nm thick SiO$_2$ (silicon dioxide) layer and a PMMA (poly methyl methacrylate) layer [see Fig. \ref{fig:WaveguideMode}(a)]. 
Figures \ref{fig:WaveguideMode}(c) and \ref{fig:WaveguideMode}(d) show the numerically calculated effective indexes and propagation lengths of waveguide modes supported by the configuration as a function of the PMMA thickness at the telecommunication wavelength of $\lambda=1550$\,nm. In the calculations, performed using the commercially available finite element software Comsol Multiphysics, the refractive index of SiO$_2$ and PMMA is assumed to be 1.45 and 1.49, respectively, while the value for gold is $0.52+i10.7$ as found from interpolation of experimental values \cite{johnson}. It is seen that the first transverse magnetic (TM) mode persists for all PMMA thicknesses, but being a surface plasmon polariton mode (with the maximum electric field at the glass-gold interface) it also features a relatively low propagation length. In order to extend the distance of which information can be carried, we choose to couple light to the first transverse electric (TE) mode, which is a photonic mode with the maximum electric field appearing away from the metal interface [Fig. \ref{fig:WaveguideMode}(b)]. It is clear that one can achieve propagation lengths of several hundreds of micrometers by a properly thick PMMA thickness. The simultaneous increase in the real part of the effective refractive index, however, signifies the need for an increasingly smaller grating period in order to reach the phase matching condition, thus potentially leading to feature sizes of the metacoupler that prevents the incorporation of a proper linear phase gradient. In order to avoid such problems, while still having a waveguide mode that is reasonably confined to the PMMA layer, we choose a PMMA thickness of 400\,nm corresponding to a TE1 mode with an effective index of $1.10$ and a propagation length of $\simeq 130$\,$\mu$m. The associated metacoupler should then feature a grating period of $\Lambda_g=\lambda/1.10\simeq 1.41$\,$\mu$m in order to couple normal incident light to the TE1 mode. 

\begin{figure}[tb]
	\centering
		\includegraphics[width=8.0cm]{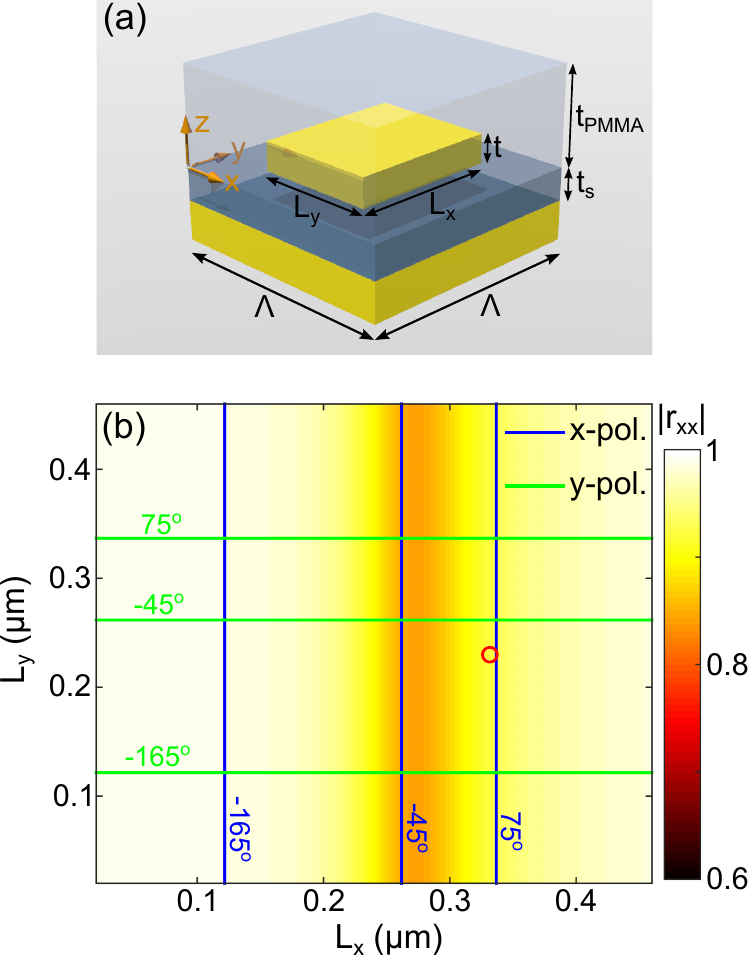}
	\caption{Design of metacouplers. (a) Sketch of unit cell of GSP-based metacoupler. (b) Calculated reflection coefficient as a function of nanobrick widths ($L_x,L_y$) for $x$-polarized normal incident light and geometrical parameters $\Lambda=470$\,nm, $t_s=70$\,nm, $t_\mathrm{PMMA}=400$\,nm, $t=50$\,nm, and wavelength $\lambda=800$\,nm. The color map shows the reflection amplitude, whereas the solid lines represent contour lines of the reflection phase for both $x$- and $y$-polarized light.}
	\label{fig:Reflection}
\end{figure}

Having clarified the waveguide configuration, we next discuss the procedure of designing the GSP-based metacouplers. The basic unit cell is schematically shown in Fig. \ref{fig:Reflection}(a), which is fundamentally the waveguide configuration with a gold nanobrick positioned atop of the SiO$_2$ layer, hereby ensuring the possibility of controlling the phase of the scattered light by utilizing nanobrick dimensions in the neighborhood of the resonant GSP configuration. The (approximate) linear phase gradient of the metacouplers is in this work achieved by incorporating three unit cells within each grating period, with adjacent unit cells featuring a difference in reflection phase of 120$^\circ$.  
In order to find the appropriate nanobrick dimensions, we perform full-wave numerical calculations of the interaction of normal incident $x$- and $y$-polarized light with the array of unit cells in Fig. \ref{fig:Reflection}(a) when the geometrical parameters take on the values $\Lambda=\Lambda_g/3=470$\,nm, $t_s=70$\,nm, $t_\mathrm{PMMA}=400$\,nm, and $t=50$\,nm. The key parameter is the complex reflection coefficient as a function of nanobricks widths ($L_x,L_y$), which is displayed in Fig. \ref{fig:Reflection}(b) for $x$-polarized light, with superimposition of phase contour lines in steps of $120^\circ$ for $y$-polarized light as well. It is seen that the metasurface is highly reflecting for most configurations. However, near $L_x=275$\,nm (keeping $L_y$ constant) the reflection amplitude features a noticeable dip accompanied by a significant change in the reflection phase. This is the signature of the GSP resonance and, together with the assumption of negligible coupling between neighboring nanobricks, the necessary ingredient in designing phase-gradient (i.e., inhomogeneous) metasurfaces.

\begin{figure*}[tb]
	\centering
		\includegraphics[width=15.0cm]{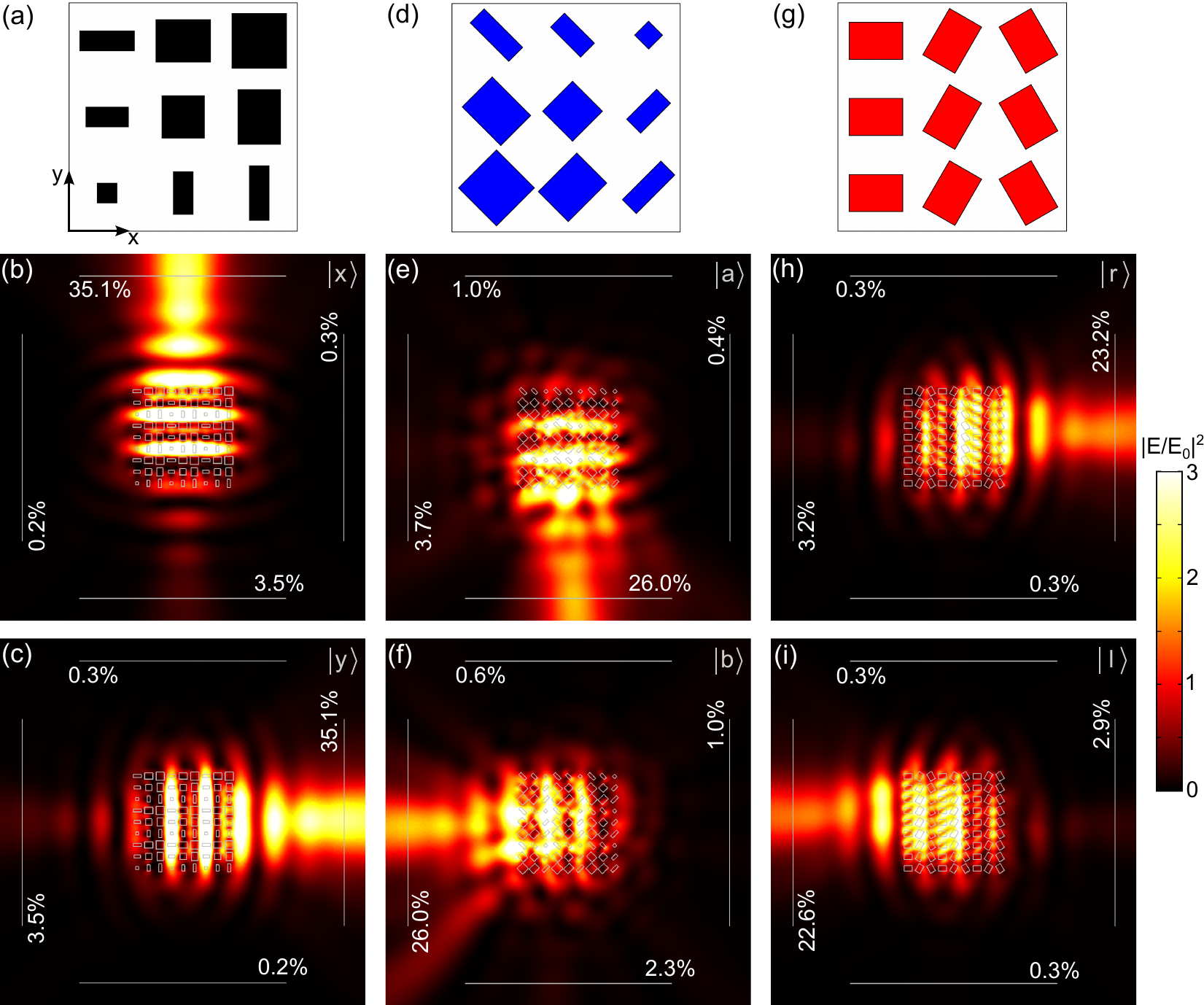}
	\caption{Performance of the individual metacouplers. (a,d,g) Top-view of super cell of coupler 1, 2, and 3, respectively. (b,c,e,f,h,i) Colormap of the intensity in the center of the PMMA layer for couplers consisting of $3\times3$ super cells when the incident light is a Gaussian beam with beam radius 3\,$\mu$m. Note that the scale bar is chosen to better highlight weak intensity features, while the numbers (in percent) displayed in the panels correspond to coupling efficiencies through the areas marked by gray lines. (b,c) Coupler 1 for incident polarization state $|x\rangle$ and $|y\rangle$; (e,f) Coupler 2 for incident polarization state $|a\rangle$ and $|b\rangle$; (h,i) Coupler 3 for incident polarization state $|r\rangle$ and $|l\rangle$.}
	\label{fig:Individual}
\end{figure*}

In the design of a unidirectional and polarization-controlled waveguide coupler for the ($\hat{\mathbf{x}},\hat{\mathbf{y}}$)-basis, which we denote metacoupler 1, we follow the previously developed approach \cite{Pors_2014}, where the $\Lambda_g \times \Lambda_g$ super cell is populated with nine nanobricks defined by the intersection of phase contour lines in Fig. \ref{fig:Reflection}(b). A top-view of the super cell is displayed in Fig. \ref{fig:Individual}(a), where the nanobricks are arranged in such a way that $x(y)$-polarized incident light experiences a phase-gradient in the $y(x)$-direction, thus ensuring unidirectional excitation of the TE1 mode.
As a way of probing the functionality of the designed super cell, we perform full-wave simulations of a coupler consisting of $3\times3$ super cells, with the incident light being a Gaussian beam with beam radius of $3$\,$\mu$m. The resulting intensity distribution in the center of the PMMA layer is shown in Figs. \ref{fig:Individual}(b) and \ref{fig:Individual}(c) for polarization states $|x\rangle$ and $|y\rangle$, which verifies that the TE1 mode is dominantly launched in the $+y$- and $+x$-direction, respectively, as a consequence of the incorporated birefringent phase gradient in the metacoupler. Moreover, the coupling efficiency, as defined by the power carried by the TE1 mode in the desired propagation direction relative to the incident power, is quite high, reaching in this numerical example $\sim 35$\% despite the fact that no attempt has been made in reaching efficient coupling. 

The second waveguide metacoupler is intended to show markedly different directional excitation of the TE1 mode for the polarization states $|a\rangle$ and $|b\rangle$. As a simple way to realize this functionality, we reuse the super cell of metacoupler 1, though this time the nanobricks are rotated 45$^\circ$ around their center of mass in the $xy$-plane, followed by an overall 180$^\circ$ rotation of the super cell [Fig. \ref{fig:Individual}(d)]. The latter rotation is implemented in order to achieve dominant launching of the TE1 mode in the $-y$- and $-x$-direction for $a$- and $b$-polarized light, respectively. This fact is evidenced in Figs. \ref{fig:Individual}(e) and \ref{fig:Individual}(f), where $\sim 26$\% of the incident power is coupled to the TE1 mode in the desired direction, hence verifying the unidirectional and birefringent response of this waveguide metacoupler. In passing we note that $b$-polarization also launches a (weaker) mode propagating towards the bottom-left corner [see Fig. \ref{fig:Individual}(f)], which ultimately is a result arising from the finite size of the metacoupler (i.e., small number of periods), hence providing phase-matching to a wider span of in-plane wave numbers (also edge effects may play an important role in launching of the TE1 waveguide mode). 

The third (and final) waveguide metacoupler should feature a birefringent response so that the TE1 mode is unidirectionally launched in different directions for the circular polarization states $|r\rangle$ and $|l\rangle$. Here, we realize such a property by implementing a geometric phase gradient, also known as the Pancharatnam-Berry phase \cite{pancharatnam_1956,berry_1984}, within the period of the grating. As discussed in detail elsewhere \cite{Luo_2015}, in order to ensure that all of the circularly polarized incident light feels the geometric phase, the basic unit cell (constituting the super cell) must operate as a half-wave plate, meaning that the reflection coefficient for $x$- and $y$-polarized light should have the same amplitude but a phase difference of 180$^\circ$. The nanobrick dimension satisfying this requirement is marked by a red circle in Fig. \ref{fig:Reflection}(b), and the corresponding three super cells are shown in Fig. \ref{fig:Individual}(g), where the neighboring nanobricks (of identical dimensions) along the $x$-axis are rotated by $60^\circ$ with respect to each other. The unidirectional and spin-dependent launching of the TE1 mode is probed through full-wave simulations [Figs. \ref{fig:Individual}(h) and \ref{fig:Individual}(i)], where it is readily seen that the mode is efficiently (i.e., coupling efficiency of $\sim 23$\%) launched along either $\pm x$-axis depending on the handedness of the incident wave.

\section{Performance of the in-plane polarimeter}
\begin{figure*}[htb]
	\centering
		\includegraphics[width=15.0cm]{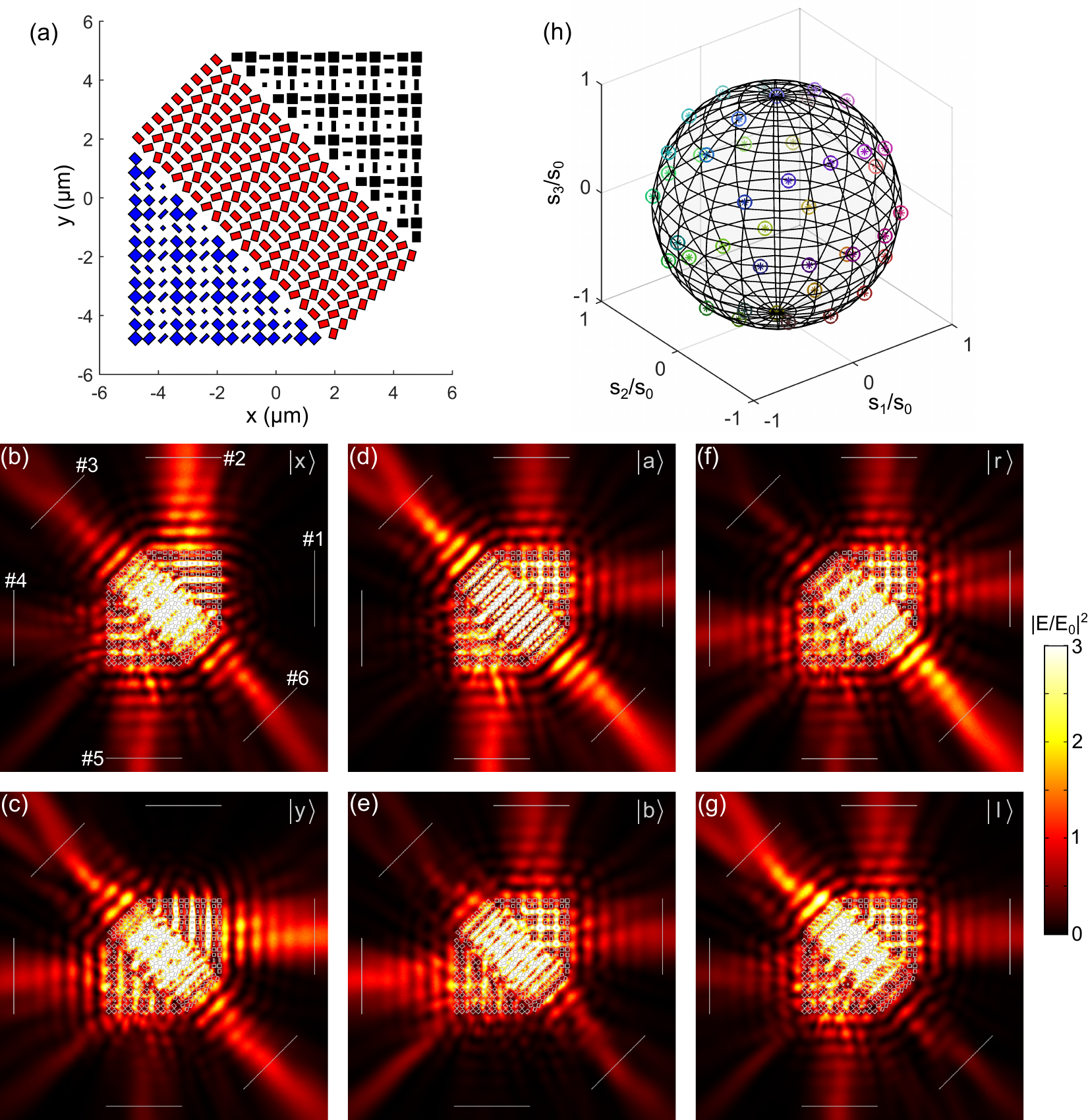}
	\caption{Performance of the combined metacoupler. (a) Top-view of the combined waveguide metacoupler. (b-g) Colormap of the intensity in the center of the PMMA layer for the metacoupler in a) when the incident light is a Gaussian beam with the beam radius of 6\,$\mu$m. The polarization state $|u\rangle$ of the beam is displayed in the upper right corner. Note that the scale bar is chosen to better highlight weak intensity features, while panel b) shows the numbering of the six ports marked with gray lines. (h) Circles and asterisks indicate retrieved and exact polarization states of the incident beam for 42 different SOPs, plotted in the ($s_1,s_2,s_3$)-space together with the Poincar{\'e} sphere. Colors are used as an aid to differentiate between the different circle-asterisk pairs.}
	\label{fig:Combined}
\end{figure*}

The previous section outlines the design of three waveguide metacouplers that each launch the TE1 modes travelling primarily along two directions, with the maximum contrast occurring for the polarization states ($|x\rangle$,$|y\rangle$), ($|a\rangle$,$|b\rangle$), and ($|r\rangle$,$|l\rangle$), respectively. Bearing in mind this design and the expressions for $s_1-s_3$ [Eqs. \eqref{eq:s1}-\eqref{eq:s3}], we can construct an all-polarization sensitive waveguide metacoupler. In this work, we incorporate the three metacouplers into one large hexagonal configuration of $\sim 100$\,$\mu$m$^2$ in footprint [Fig. \ref{fig:Combined}(a)]. 
The exact size of the waveguide metacoupler is not a critical parameter, but in order to avoid too divergent TE1 beams we have ensured that each side of the hexagon is considerably larger than the wavelength. This has the consequence that metacoupler 3 occupies $\sim46\%$ of the area and, hence, may direct more power in those directions. Noting, however, that the polarization-sensitivity is not dependent on the absolute coupling efficiencies (which, for example, can be changed by the width of the incident beam), it is clear that the proposed procedure should still work.

As a way to illustrate the all-polarization sensitivity of the combined waveguide metacoupler, we display the intensity in the PMMA layer in an area of $30\times 30$\,$\mu$m$^2$ for the six extreme polarizations [Figs. \ref{fig:Combined}(b)-\ref{fig:Combined}(g)], while the calculated coupling efficiencies in the six launching directions, as evaluated at the ports marked in Fig. \ref{fig:Combined}(b), are presented in Table \ref{tab:Efficiency}. 
\begin{table}[tb]
	\centering
		\begin{tabular}{c|cccccc}
		\hline
			Port \# & $|x\rangle$ & $|y\rangle$ & $|a\rangle$ & $|b\rangle$ & $|r\rangle$ & $|l\rangle$ \\ \hline 
			1 & 0.1 & 5.8 & 2.9 & 3.0 & 3.0 & 2.9 \\
			2 & 5.6 & 0.1 & 2.8 & 2.8 & 2.6 & 3.0 \\
			3 & 2.4 & 3.0 & 3.8 & 1.5 & 1.0 & 4.4 \\
			4 & 1.3 & 3.3 & 0.4 & 4.2 & 2.7 & 1.9 \\
			5 & 3.0 & 1.4 & 4.0 & 0.3 & 2.3 & 2.1 \\
			6 & 2.8 & 3.4 & 4.4 & 1.7 & 4.9 & 1.2 \\ \hline
			Sum & 15.2 & 17.0 & 18.3 & 13.5 & 14.2 & 15.5 \\ \hline
			$D_1$ ($s_1/s_0$) & 0.97 (1)&-0.98 (-1) & -0.02 (0) & -0.04 (0) & -0.07 (0) & 0.01 (0) \\
			$D_2$ ($s_2/s_0$) & 0.39 (0)& -0.42 (0) & 0.80 (1) & -0.86 (-1) & -0.09 (0) & 0.05 (0) \\
			$D_3$ ($s_3/s_0$) & 0.07 (0)& 0.06 (0) & 0.06 (0) & 0.07 (0) & 0.67 (1) & -0.58 (-1) \\ \hline
		\end{tabular}
	\caption{Coupling efficiencies for the combined metacoupler. Rows 2-7 display the coupling efficiencies (in percentage) between the incident Gaussian beam (with beam radius 6\,$\mu$m) of polarization $|u\rangle$ and the TE1 mode, evaluated at the six ports defined in Fig. \ref{fig:Combined}(b). The 8th row shows the total coupling efficiency. The three bottom rows display the normalized contrast in coupling efficiency for the launching directions specified by the three metacouplers, with numbers in parenthesis corresponding to the normalized Stokes parameters.}
	\label{tab:Efficiency}
\end{table}
It transpires that the combined metacoupler launches the TE1 modes propagating mainly in the six designed directions, with the power distribution in the six channels being strongly polarization dependent. For example, it is seen, depending on the SOP, that one of the six channels is suppressed. 
In fact, the normalized contrast in launching efficiency for each of the three metacouplers, denoted $D_1-D_3$ [and corresponding to the difference in power flow through the three pairs of ports (\#1,\#2), (\#4,\#5) and (\#3,\#6) normalized by their respective sum], does mimic the behavior of the Stokes parameters $s_1-s_3$ on the SOP, as evident from Table \ref{tab:Efficiency}. The largest discrepancy is found for metacoupler 2 ($D_2$), which is somewhat expected as the design is directly derived from metacoupler 1 without any further optimization. Despite the apparent convenience in having only three parameters for the description of the SOP, we emphasize that, unlike related work \cite{Pors_2015}, there is no mathematical equivalence between $D_1-D_3$ and $s_1/s_0-s_3/s_0$, nor is it even possible to find a linear relation (ie., device matrix) between those quantities that is valid for all SOPs. The reason for the incompatibility in using $D_1-D_3$ for the retrieval of the SOP owes to the fact that the incident light launches only TE-polarized modes, meaning that the relative coupling efficiencies between all the six propagation directions do carry information about the SOP. This fact is conveniently illustrated for the polarization states $|a\rangle$ and $|b\rangle$ [Figs. \ref{fig:Combined}(d) and \ref{fig:Combined}(e)], which feature practically the same $D_3$ value, but the launching efficiency of metacoupler 3 (port \#3 and \#6) is considerably higher for $|a\rangle$ due to the optimal orientation relative to the SOP. The best performance of the polarimeter can only be achieved by properly relating the six coupling efficiencies to the three Stokes parameters. If $\mathbf{S}$ and $\mathbf{C}$ denote $3\times 1$ and $6\times 1$ vectors containing the normalized Stokes parameters and coupling efficiencies, respectively, the $3\times6$ device matrix $\mathbf{A}$ should (ideally) satisfy the relation
\begin{equation}
\mathbf{S}=\mathbf{A}\mathbf{C},
\label{eq:DeviceMatrix}
\end{equation}
for any SOP, meaning that from the calculated/measured coupling efficiencies one can immediately obtain knowledge of the SOP. In general, $\mathbf{A}$ must be obtained from a calibration procedure that preferably includes polarization states covering all octants of the Poincar{\'e} sphere, hereby allowing us to find $\mathbf{A}$ in a least-squares sense: $\mathbf{A}=\tilde{\mathbf{S}}\tilde{\mathbf{C}}^+$, where $\tilde{\mathbf{S}}$ and $\tilde{\mathbf{C}}$ are $3\times N$ and $6\times N$ matrices, respectively, $N$ is the number of calibration points, and $^+$ denotes the pseudo-inverse. In order to test the accuracy of the linear relationship in Eq. \eqref{eq:DeviceMatrix}, we have retrieved $\mathbf{A}$ from a calibration set containing $N=42$ different polarizations that cover the whole Poincar{\'e} sphere, as illustrated by asterisks in Fig. \ref{fig:Combined}(h). Here,
\begin{equation}
\mathbf{A}=\begin{pmatrix}
-16.7 & 17.5 & -4.0 & 2.2 & 5.6 & -2.1 \\
3.1 & -5.7 & 12.2 & -34.2 & -0.6 & 17.9 \\
-12.4 & -11.6 & -26.1 & 14.6 & 13.3 & 25.0 	
\end{pmatrix},
\label{eq:A}
\end{equation}  
but it should be emphasized that the exact values of $\mathbf{A}$ are, in the considered situation, configuration dependent, since they depend on the size of the combined metacoupler and the width of the incident beam. This limitation can, however, be circumvented by considering a plane-wave incidence, i.e., by considering a metacoupler much smaller than the incident beam. As a way to visualize the accuracy of the suggested procedure, the coordinates defined by the columns of $\mathbf{A}\tilde{\mathbf{C}}$ are displayed as circular markers in Fig. \ref{fig:Combined}(h), hereby demonstrating a perfect overlap with the Stokes parameters. The high accuracy of Eq. \eqref{eq:DeviceMatrix} is also confirmed by the 2-norm error $||\mathbf{S}-\mathbf{A}\mathbf{C}||_2$, which is of the order of $10^{-7}$ for all 42 SOPs. We note that these small errors, corresponding to determining the Stokes parameters with an accuracy of six decimals, are obtained using the numerically calculated coupling efficiencies with full precision [i.e., not the rounded off data presented in Table \ref{tab:Efficiency} and Eq. \eqref{eq:A}], hereby highlighting the perfect linear relationship between coupling efficiencies and the Stokes parameters. In realistic scenarios, however, multiple sources of noise will lower the accuracy with which the SOP can be determined and, hence, the resolution at which different SOPs can be distinguished \cite{Foreman_2010}. Without going into detailed discussion of certain types of noise distributions (e.g., Poisson and Gaussian) and their influence on the polarization resolution, we notice that the 2-norm condition number of Eq. \eqref{eq:A}, which describes the sensitivity of the linear system to noise \cite{Twietmeyer_2008}, is $\simeq 2$. Since this value is close to the condition number of device matrices from other polarimeter designs ($\mathrm{cond}(A)\sim 1.7-2.9$ \cite{Lara_2009}), we believe that the proposed in-plane polarimeter is similar to others with respect to the detrimental influence of noise.

As a final comment to the above discussion, it should be noted that most polarimeter designs utilize the linear relation $\mathbf{C}=\mathbf{B}\mathbf{S}_{4}$, where the four-vector $\mathbf{S}_4=\left[s_0,s_1,s_2,s_3\right]^T$ is treated as the input signal, $\mathbf{B}$ is the device matrix, and $\mathbf{C}$ is a vector containing the detected signals \cite{Foreman_2010,Twietmeyer_2008,Lara_2009,Azzam_1988}. In relation to Eq. \eqref{eq:DeviceMatrix}, several things are worth noting. First, we have throughout this work focused on the determination of the SOP through the retrieval of $s_1-s_3$, excluding any considerations of $s_0$ that is just used as a normalizing parameter. The proposed polarimeter, however, can easily be extended to provide information on all elements of $\mathbf{S}_4$ by extending $\mathbf{A}$ to a $4\times 6$ device matrix, where the upper row (in normalized units) takes on the values $\left[26.1, 23.5, -7.0, -6.2, 0.9, -3.9 \right]$, while the latter three rows are identical to Eq. \eqref{eq:A}. Secondly, for conventional polarimeters the unknown Stokes vector is retrieved by the relation $\mathbf{S}_4=\mathbf{B}^+\mathbf{C}$, thus implying the need to invert the device matrix. Since the matrix (pseudo) inverse can only be computed in a least-squares sense for non-square matrices, we have preferred to use the matrix system of Eq. \eqref{eq:DeviceMatrix}. Finally, it is worth noting that the preceding discussion has exclusively considered fully polarized light (i.e., $s_1^2+s_2^2+s_3^2=s_0^2$), though the proposed polarimeter can also handle partially polarized light (i.e., $s_1^2+s_2^2+s_3^2<s_0^2$), as seen by the fact that the (time-averaged) diffraction contrasts will decrease as the degree of polarization decreases. In the extreme case of unpolarized light, the six coupling efficiencies are the average of the result from two orthogonal polarizations, hereby ensuring that the Stokes parameters $s_1-s_3$ will be zero.

\section{Conclusion}
In summary, we have designed a compact in-plane polarimeter that couples incident light into waveguide modes propagating along six different directions, with the coupling efficiencies being dictated by the SOP. This allows one to realize simultaneous detection of the Stokes parameters. The functionality and high accuracy of the proposed device has been verified numerically by performing full-wave calculations of a $\sim 100$\,$\mu$m$^2$ device at the wavelength of $1.55$\,$\mu$m. The polarimeter is based on three GSP-based birefringent metasurfaces that each features a linear phase-gradient that is dependent on the SOP, thus ensuring unidirectional and all-polarization sensitive excitation of the waveguide modes.
 
We note that the choice of the design wavelength at $1.55$\,$\mu$m is merely to illustrate its potential usage in compact integrated optical circuitry, but the design strategy can be transferred to any frequency range of interest, being it either at optical wavelengths \cite{Huang_2013} or the microwave regime \cite{Sun_2012}. Moreover, the losses associated with plasmonic metasurfaces, which in our case study amounts to $\sim 35$\% of the incident power, can be redeemed by utilizing high-dielectric nanostructures instead \cite{Yang_2014}. Regarding the spectral bandwidth of the proposed design, it should be noted that phase-matching with the TE1 mode is achieved through grating coupling, which makes the polarimeter inherently narrow-band since the period of the grating must be close to the wavelength of the mode. As such, one must design the polarimeter to the wavelength of operation. Also, it is worth noting that conventional polarimeters typically measure the SOP in a destructive (i.e., strongly modifying or extinction of the incident beam) or perturbative way. Since our polarimeter is based on an opaque metal film, thus preventing any light to be transmitted, it operates in a destructive manner. At the same time, the perturbative regime can always be approached by utilizing a beam-splitter in front of a polarimeter performing a destructive measurement.

Finally, we would like to stress that the suggested in-plane polarimeter can be realized by only one step of electron-beam lithography, while simple proof-of-concept experiments can be performed by placing out-coupling gratings along the six in-plane propagation directions, with the associated scattered light being a measure of the coupling efficiencies. Moreover, we foresee the possibility of a compact circuitry with built-in plasmonic detectors that are integrated into spatially-confined waveguides \cite{Neutens_2013,Zhanghua_2015}.

\begin{acknowledgments}
We acknowledge financial support for this work from the European Research Council, grant 341054 (PLAQNAP), and the University of Southern Denmark (SDU2020 funding). We also acknowledge support from the DeIC National HPC Center for access to computational resources on Abacus 2.0. 
\end{acknowledgments}


\begin{thebibliography}{34}%
\makeatletter
\providecommand \@ifxundefined [1]{%
 \@ifx{#1\undefined}
}%
\providecommand \@ifnum [1]{%
 \ifnum #1\expandafter \@firstoftwo
 \else \expandafter \@secondoftwo
 \fi
}%
\providecommand \@ifx [1]{%
 \ifx #1\expandafter \@firstoftwo
 \else \expandafter \@secondoftwo
 \fi
}%
\providecommand \natexlab [1]{#1}%
\providecommand \enquote  [1]{``#1''}%
\providecommand \bibnamefont  [1]{#1}%
\providecommand \bibfnamefont [1]{#1}%
\providecommand \citenamefont [1]{#1}%
\providecommand \href@noop [0]{\@secondoftwo}%
\providecommand \href [0]{\begingroup \@sanitize@url \@href}%
\providecommand \@href[1]{\@@startlink{#1}\@@href}%
\providecommand \@@href[1]{\endgroup#1\@@endlink}%
\providecommand \@sanitize@url [0]{\catcode `\\12\catcode `\$12\catcode
  `\&12\catcode `\#12\catcode `\^12\catcode `\_12\catcode `\%12\relax}%
\providecommand \@@startlink[1]{}%
\providecommand \@@endlink[0]{}%
\providecommand \url  [0]{\begingroup\@sanitize@url \@url }%
\providecommand \@url [1]{\endgroup\@href {#1}{\urlprefix }}%
\providecommand \urlprefix  [0]{URL }%
\providecommand \Eprint [0]{\href }%
\providecommand \doibase [0]{http://dx.doi.org/}%
\providecommand \selectlanguage [0]{\@gobble}%
\providecommand \bibinfo  [0]{\@secondoftwo}%
\providecommand \bibfield  [0]{\@secondoftwo}%
\providecommand \translation [1]{[#1]}%
\providecommand \BibitemOpen [0]{}%
\providecommand \bibitemStop [0]{}%
\providecommand \bibitemNoStop [0]{.\EOS\space}%
\providecommand \EOS [0]{\spacefactor3000\relax}%
\providecommand \BibitemShut  [1]{\csname bibitem#1\endcsname}%
\let\auto@bib@innerbib\@empty
\bibitem [{\citenamefont {Tinbergen}(1996)}]{Tinbergen_1996}%
  \BibitemOpen
  \bibfield  {author} {\bibinfo {author} {\bibfnamefont {J.}\ \bibnamefont
  {Tinbergen}},\ }\href@noop {} {\emph {\bibinfo {title} {Astronomical
  Polarimetry}}}\ (\bibinfo  {publisher} {Cambridge University Press},\
  \bibinfo {address} {New York},\ \bibinfo {year} {1996})\BibitemShut {NoStop}%
\bibitem [{\citenamefont {Collings}\ and\ \citenamefont
  {Caruso}(1997)}]{Collings_1997}%
  \BibitemOpen
  \bibfield  {author} {\bibinfo {author} {\bibfnamefont {A.~F.}\ \bibnamefont
  {Collings}}\ and\ \bibinfo {author} {\bibfnamefont {F.}\ \bibnamefont
  {Caruso}},\ }\bibfield  {title} {\enquote {\bibinfo {title} {Biosensors:
  recent advances},}\ }\href@noop {} {\bibfield  {journal} {\bibinfo  {journal}
  {Rep. Prog. Phys.}\ }\textbf {\bibinfo {volume} {60}},\ \bibinfo {pages}
  {1397} (\bibinfo {year} {1997})}\BibitemShut {NoStop}%
\bibitem [{\citenamefont {Tyo}\ \emph {et~al.}(2006)\citenamefont {Tyo},
  \citenamefont {Goldstein}, \citenamefont {Chenault},\ and\ \citenamefont
  {Shaw}}]{Tyo_2006}%
  \BibitemOpen
  \bibfield  {author} {\bibinfo {author} {\bibfnamefont {J.~S.}\
  \bibnamefont {Tyo}}, \bibinfo {author} {\bibfnamefont {D.~L.}\
  \bibnamefont {Goldstein}}, \bibinfo {author} {\bibfnamefont {D.~B.}\
  \bibnamefont {Chenault}}, \ and\ \bibinfo {author} {\bibfnamefont
  {J.~A.}\ \bibnamefont {Shaw}},\ }\bibfield  {title} {\enquote {\bibinfo
  {title} {Review of passive imaging polarimetry for remote sensing
  applications},}\ }\href {\doibase 10.1364/AO.45.005453} {\bibfield  {journal}
  {\bibinfo  {journal} {Appl. Opt.}\ }\textbf {\bibinfo {volume} {45}},\
  \bibinfo {pages} {5453} (\bibinfo {year} {2006})}\BibitemShut {NoStop}%
\bibitem [{\citenamefont {Lepetit}\ and\ \citenamefont
  {Kant{\'e}}(2015)}]{Lepetit_2015}%
  \BibitemOpen
  \bibfield  {author} {\bibinfo {author} {\bibfnamefont {T.}\ \bibnamefont
  {Lepetit}}\ and\ \bibinfo {author} {\bibfnamefont {B.}\ \bibnamefont
  {Kant{\'e}}},\ }\bibfield  {title} {\enquote {\bibinfo {title} {Metasurfaces:
  Simultaneous stokes parameters},}\ }\href@noop {} {\bibfield  {journal}
  {\bibinfo  {journal} {Nat. Photon.}\ }\textbf {\bibinfo {volume} {9}},\
  \bibinfo {pages} {709} (\bibinfo {year} {2015})}\BibitemShut {NoStop}%
\bibitem [{\citenamefont {Bohren}\ and\ \citenamefont
  {Huffman}(2004)}]{bohren}%
  \BibitemOpen
  \bibfield  {author} {\bibinfo {author} {\bibfnamefont {C.~F.}\
  \bibnamefont {Bohren}}\ and\ \bibinfo {author} {\bibfnamefont {D.~R.}\
  \bibnamefont {Huffman}},\ }\href@noop {} {\emph {\bibinfo {title} {Absorption
  and Scattering of Light by Small Particles}}}\ (\bibinfo  {publisher}
  {Wiley-VCH Verlag GmbH},\ \bibinfo {address} {Weinheim},\ \bibinfo {year}
  {2004})\BibitemShut {NoStop}%
\bibitem [{\citenamefont {Compain}\ and\ \citenamefont
  {Drevillon}(1998)}]{Compain_1998}%
  \BibitemOpen
  \bibfield  {author} {\bibinfo {author} {\bibfnamefont {E.}\ \bibnamefont
  {Compain}}\ and\ \bibinfo {author} {\bibfnamefont {B.}\ \bibnamefont
  {Drevillon}},\ }\bibfield  {title} {\enquote {\bibinfo {title} {Broadband
  division-of-amplitude polarimeter based on uncoated prisms},}\ }\href
  {\doibase 10.1364/AO.37.005938} {\bibfield  {journal} {\bibinfo  {journal}
  {Appl. Opt.}\ }\textbf {\bibinfo {volume} {37}},\ \bibinfo {pages}
  {5938} (\bibinfo {year} {1998})}\BibitemShut {NoStop}%
\bibitem [{\citenamefont {Yu}\ and\ \citenamefont {Capasso}(2014)}]{Yu_2014}%
  \BibitemOpen
  \bibfield  {author} {\bibinfo {author} {\bibfnamefont {N.}~\bibnamefont
  {Yu}}\ and\ \bibinfo {author} {\bibfnamefont {F.}~\bibnamefont {Capasso}},\
  }\bibfield  {title} {\enquote {\bibinfo {title} {Flat optics with designer
  metasurfaces},}\ }\href@noop {} {\bibfield  {journal} {\bibinfo  {journal}
  {Nat. Mater.}\ }\textbf {\bibinfo {volume} {13}},\ \bibinfo {pages}
  {139} (\bibinfo {year} {2014})}\BibitemShut {NoStop}%
\bibitem [{\citenamefont {Estakhri}\ and\ \citenamefont
  {Al\`{u}}(2016)}]{Estakhri_2016}%
  \BibitemOpen
  \bibfield  {author} {\bibinfo {author} {\bibfnamefont {N.~M.}\
  \bibnamefont {Estakhri}}\ and\ \bibinfo {author} {\bibfnamefont {A.}\
  \bibnamefont {Al\`{u}}},\ }\bibfield  {title} {\enquote {\bibinfo {title}
  {Recent progress in gradient metasurfaces},}\ }\href {\doibase
  10.1364/JOSAB.33.000A21} {\bibfield  {journal} {\bibinfo  {journal} {J. Opt.
  Soc. Am. B}\ }\textbf {\bibinfo {volume} {33}},\ \bibinfo {pages} {A21}
  (\bibinfo {year} {2016})}\BibitemShut {NoStop}%
\bibitem [{\citenamefont {Gori}(1999)}]{Gori_99}%
  \BibitemOpen
  \bibfield  {author} {\bibinfo {author} {\bibfnamefont {F.}\ \bibnamefont
  {Gori}},\ }\bibfield  {title} {\enquote {\bibinfo {title} {Measuring stokes
  parameters by means of a polarization grating},}\ }\href@noop {} {\bibfield
  {journal} {\bibinfo  {journal} {Opt. Lett.}\ }\textbf {\bibinfo {volume}
  {24}},\ \bibinfo {pages} {584} (\bibinfo {year} {1999})}\BibitemShut
  {NoStop}%
\bibitem [{\citenamefont {Bomzon}\ \emph {et~al.}(2001)\citenamefont {Bomzon},
  \citenamefont {Biener}, \citenamefont {Kleiner},\ and\ \citenamefont
  {Hasman}}]{Bomzon_01}%
  \BibitemOpen
  \bibfield  {author} {\bibinfo {author} {\bibfnamefont {Z.}\ \bibnamefont
  {Bomzon}}, \bibinfo {author} {\bibfnamefont {G.}\ \bibnamefont
  {Biener}}, \bibinfo {author} {\bibfnamefont {V.}\ \bibnamefont
  {Kleiner}}, \ and\ \bibinfo {author} {\bibfnamefont {E.}\ \bibnamefont
  {Hasman}},\ }\bibfield  {title} {\enquote {\bibinfo {title} {Spatial
  fourier-transform polarimetry using space-variant subwavelength metal-stripe
  polarizers},}\ }\href@noop {} {\bibfield  {journal} {\bibinfo  {journal}
  {Opt. Lett.}\ }\textbf {\bibinfo {volume} {26}},\ \bibinfo {pages}
  {1711} (\bibinfo {year} {2001})}\BibitemShut {NoStop}%
\bibitem [{\citenamefont {Afshinmanesh}\ \emph {et~al.}(2012)\citenamefont
  {Afshinmanesh}, \citenamefont {White}, \citenamefont {Cai},\ and\
  \citenamefont {Brongersma}}]{Afshinmanesh_2012}%
  \BibitemOpen
  \bibfield  {author} {\bibinfo {author} {\bibfnamefont {F.}\
  \bibnamefont {Afshinmanesh}}, \bibinfo {author} {\bibfnamefont {J.~S.}\
  \bibnamefont {White}}, \bibinfo {author} {\bibfnamefont {W.}\
  \bibnamefont {Cai}}, \ and\ \bibinfo {author} {\bibfnamefont {M.~L.}\
  \bibnamefont {Brongersma}},\ }\bibfield  {title} {\enquote {\bibinfo {title}
  {Measurement of the polarization state of light using an integrated plasmonic
  polarimeter},}\ }\href@noop {} {\bibfield  {journal} {\bibinfo  {journal}
  {Nanophotonics}\ }\textbf {\bibinfo {volume} {1}},\ \bibinfo {pages}
  {125} (\bibinfo {year} {2012})}\BibitemShut {NoStop}%
\bibitem [{\citenamefont {Mandel}\ \emph {et~al.}(2013)\citenamefont {Mandel},
  \citenamefont {Gollub}, \citenamefont {Bendoym},\ and\ \citenamefont
  {Crouse}}]{Mandel_2013}%
  \BibitemOpen
  \bibfield  {author} {\bibinfo {author} {\bibfnamefont {I.}~\bibnamefont
  {Mandel}}, \bibinfo {author} {\bibfnamefont {J.~N.}\ \bibnamefont {Gollub}},
  \bibinfo {author} {\bibfnamefont {I.}~\bibnamefont {Bendoym}}, \ and\
  \bibinfo {author} {\bibfnamefont {D.~T.}\ \bibnamefont {Crouse}},\ }\bibfield
   {title} {\enquote {\bibinfo {title} {Theory and design of a novel integrated
  polarimetric sensor utilizing a light sorting metamaterial grating},}\ }\href
  {\doibase 10.1109/JSEN.2012.2220539} {\bibfield  {journal} {\bibinfo
  {journal} {IEEE Sens. J.}\ }\textbf {\bibinfo {volume} {13}},\ \bibinfo
  {pages} {618} (\bibinfo {year} {2013})}\BibitemShut {NoStop}%
\bibitem [{\citenamefont {Wen}\ \emph {et~al.}(2015)\citenamefont {Wen},
  \citenamefont {Yue}, \citenamefont {Kumar}, \citenamefont {Ma}, \citenamefont
  {Chen}, \citenamefont {Ren}, \citenamefont {Kremer}, \citenamefont
  {Gerardot}, \citenamefont {Taghizadeh}, \citenamefont {Buller},\ and\
  \citenamefont {Chen}}]{Wen_2015}%
  \BibitemOpen
  \bibfield  {author} {\bibinfo {author} {\bibfnamefont {D.}\ \bibnamefont
  {Wen}}, \bibinfo {author} {\bibfnamefont {F.}\ \bibnamefont {Yue}},
  \bibinfo {author} {\bibfnamefont {S.}\ \bibnamefont {Kumar}}, \bibinfo
  {author} {\bibfnamefont {Y.}\ \bibnamefont {Ma}}, \bibinfo {author}
  {\bibfnamefont {M.}\ \bibnamefont {Chen}}, \bibinfo {author} {\bibfnamefont
  {X.}\ \bibnamefont {Ren}}, \bibinfo {author} {\bibfnamefont {P.~E.}\
  \bibnamefont {Kremer}}, \bibinfo {author} {\bibfnamefont {B.~D.}\
  \bibnamefont {Gerardot}}, \bibinfo {author} {\bibfnamefont {M.~R.}\
  \bibnamefont {Taghizadeh}}, \bibinfo {author} {\bibfnamefont {G.~S.}\
  \bibnamefont {Buller}}, \ and\ \bibinfo {author} {\bibfnamefont {X.}\
  \bibnamefont {Chen}},\ }\bibfield  {title} {\enquote {\bibinfo {title}
  {Metasurface for characterization of the polarization state of light},}\
  }\href@noop {} {\bibfield  {journal} {\bibinfo  {journal} {Opt. Express}\
  }\textbf {\bibinfo {volume} {23}},\ \bibinfo {pages} {10272} (\bibinfo
  {year} {2015})}\BibitemShut {NoStop}%
\bibitem [{\citenamefont {Shaltout}\ \emph {et~al.}(2015)\citenamefont
  {Shaltout}, \citenamefont {Liu}, \citenamefont {Kildishev},\ and\
  \citenamefont {Shalaev}}]{Shaltout_2015}%
  \BibitemOpen
  \bibfield  {author} {\bibinfo {author} {\bibfnamefont {A.}\ \bibnamefont
  {Shaltout}}, \bibinfo {author} {\bibfnamefont {J.}\ \bibnamefont
  {Liu}}, \bibinfo {author} {\bibfnamefont {A.}\ \bibnamefont
  {Kildishev}}, \ and\ \bibinfo {author} {\bibfnamefont {V.}\
  \bibnamefont {Shalaev}},\ }\bibfield  {title} {\enquote {\bibinfo {title}
  {Photonic spin hall effect in gap-plasmon metasurfaces for on-chip
  chiroptical spectroscopy},}\ }\href {\doibase 10.1364/OPTICA.2.000860}
  {\bibfield  {journal} {\bibinfo  {journal} {Optica}\ }\textbf {\bibinfo
  {volume} {2}},\ \bibinfo {pages} {860} (\bibinfo {year}
  {2015})}\BibitemShut {NoStop}%
\bibitem [{\citenamefont {Mueller}\ \emph {et~al.}(2016)\citenamefont
  {Mueller}, \citenamefont {Leosson},\ and\ \citenamefont
  {Capasso}}]{Mueller_2016}%
  \BibitemOpen
  \bibfield  {author} {\bibinfo {author} {\bibfnamefont {J.~P.~Balthasar}\
  \bibnamefont {Mueller}}, \bibinfo {author} {\bibfnamefont {K.}\
  \bibnamefont {Leosson}}, \ and\ \bibinfo {author} {\bibfnamefont {F.}\
  \bibnamefont {Capasso}},\ }\bibfield  {title} {\enquote {\bibinfo {title}
  {Ultracompact metasurface in-line polarimeter},}\ }\href {\doibase
  10.1364/OPTICA.3.000042} {\bibfield  {journal} {\bibinfo  {journal} {Optica}\
  }\textbf {\bibinfo {volume} {3}},\ \bibinfo {pages} {42} (\bibinfo {year}
  {2016})}\BibitemShut {NoStop}%
\bibitem [{\citenamefont {Pors}\ \emph
  {et~al.}(2015{\natexlab{a}})\citenamefont {Pors}, \citenamefont {Nielsen},\
  and\ \citenamefont {Bozhevolnyi}}]{Pors_2015}%
  \BibitemOpen
  \bibfield  {author} {\bibinfo {author} {\bibfnamefont {A.}\ \bibnamefont
  {Pors}}, \bibinfo {author} {\bibfnamefont {M.~G.}\ \bibnamefont
  {Nielsen}}, \ and\ \bibinfo {author} {\bibfnamefont {S.~I.}\ \bibnamefont
  {Bozhevolnyi}},\ }\bibfield  {title} {\enquote {\bibinfo {title} {Plasmonic
  metagratings for simultaneous determination of stokes parameters},}\ }\href
  {\doibase 10.1364/OPTICA.2.000716} {\bibfield  {journal} {\bibinfo  {journal}
  {Optica}\ }\textbf {\bibinfo {volume} {2}},\ \bibinfo {pages} {716}
  (\bibinfo {year} {2015}{\natexlab{a}})}\BibitemShut {NoStop}%
\bibitem [{\citenamefont {Pozar}\ \emph {et~al.}(1997)\citenamefont {Pozar},
  \citenamefont {Targonski},\ and\ \citenamefont {Syrigos}}]{Pozar_1997}%
  \BibitemOpen
  \bibfield  {author} {\bibinfo {author} {\bibfnamefont {D.~M.}\
  \bibnamefont {Pozar}}, \bibinfo {author} {\bibfnamefont {S.~D.}\
  \bibnamefont {Targonski}}, \ and\ \bibinfo {author} {\bibfnamefont {H.D.}\
  \bibnamefont {Syrigos}},\ }\bibfield  {title} {\enquote {\bibinfo {title}
  {Design of millimeter wave microstrip reflectarrays},}\ }\href@noop {}
  {\bibfield  {journal} {\bibinfo  {journal} {IEEE Trans. Antennas Propag.}\
  }\textbf {\bibinfo {volume} {45}},\ \bibinfo {pages} {287} (\bibinfo
  {year} {1997})}\BibitemShut {NoStop}%
\bibitem [{\citenamefont {Pors}\ \emph {et~al.}(2013)\citenamefont {Pors},
  \citenamefont {Albrektsen}, \citenamefont {Radko},\ and\ \citenamefont
  {Bozhevolnyi}}]{Pors_2013}%
  \BibitemOpen
  \bibfield  {author} {\bibinfo {author} {\bibfnamefont {A.}~\bibnamefont
  {Pors}}, \bibinfo {author} {\bibfnamefont {O.}~\bibnamefont {Albrektsen}},
  \bibinfo {author} {\bibfnamefont {I.~P.}\ \bibnamefont {Radko}}, \ and\
  \bibinfo {author} {\bibfnamefont {S.~I.}\ \bibnamefont {Bozhevolnyi}},\
  }\bibfield  {title} {\enquote {\bibinfo {title} {Gap plasmon-based
  metasurfaces for total control of reflected light},}\ }\href@noop {}
  {\bibfield  {journal} {\bibinfo  {journal} {Sci. Rep.}\ }\textbf {\bibinfo
  {volume} {3}},\ \bibinfo {pages} {2155} (\bibinfo {year} {2013})}\BibitemShut
  {NoStop}%
\bibitem [{\citenamefont {Pors}\ \emph
  {et~al.}(2015{\natexlab{b}})\citenamefont {Pors}, \citenamefont {Nielsen},\
  and\ \citenamefont {Bozhevolnyi}}]{Pors_2015_2}%
  \BibitemOpen
  \bibfield  {author} {\bibinfo {author} {\bibfnamefont {A.}\ \bibnamefont
  {Pors}}, \bibinfo {author} {\bibfnamefont {M.~G.}\ \bibnamefont
  {Nielsen}}, \ and\ \bibinfo {author} {\bibfnamefont {S.~I.}\ \bibnamefont
  {Bozhevolnyi}},\ }\bibfield  {title} {\enquote {\bibinfo {title} {Analog
  computing using reflective plasmonic metasurfaces},}\ }\href {\doibase
  10.1021/nl5047297} {\bibfield  {journal} {\bibinfo  {journal} {Nano Lett.}\
  }\textbf {\bibinfo {volume} {15}},\ \bibinfo {pages} {791} (\bibinfo
  {year} {2015}{\natexlab{b}})}\BibitemShut {NoStop}%
\bibitem [{\citenamefont {Chen}\ \emph {et~al.}(2014)\citenamefont {Chen},
  \citenamefont {Yang}, \citenamefont {Wang}, \citenamefont {Huang},
  \citenamefont {Sun}, \citenamefont {Chiang}, \citenamefont {Liao},
  \citenamefont {Hsu}, \citenamefont {Lin}, \citenamefont {Sun}, \citenamefont
  {Zhou}, \citenamefont {Liu},\ and\ \citenamefont {Tsai}}]{Chen_2014}%
  \BibitemOpen
  \bibfield  {author} {\bibinfo {author} {\bibfnamefont {W.~T.}\ \bibnamefont
  {Chen}}, \bibinfo {author} {\bibfnamefont {K.-Y.}\ \bibnamefont {Yang}},
  \bibinfo {author} {\bibfnamefont {C.-M.}\ \bibnamefont {Wang}}, \bibinfo
  {author} {\bibfnamefont {Y.-W.}\ \bibnamefont {Huang}}, \bibinfo {author}
  {\bibfnamefont {G.}~\bibnamefont {Sun}}, \bibinfo {author} {\bibfnamefont
  {I-D.}\ \bibnamefont {Chiang}}, \bibinfo {author} {\bibfnamefont {C.~Y.}\
  \bibnamefont {Liao}}, \bibinfo {author} {\bibfnamefont {W.-L.}\ \bibnamefont
  {Hsu}}, \bibinfo {author} {\bibfnamefont {H.~T.}\ \bibnamefont {Lin}},
  \bibinfo {author} {\bibfnamefont {S.}~\bibnamefont {Sun}}, \bibinfo {author}
  {\bibfnamefont {L.}~\bibnamefont {Zhou}}, \bibinfo {author} {\bibfnamefont
  {A.~Q.}\ \bibnamefont {Liu}}, \ and\ \bibinfo {author} {\bibfnamefont
  {D.~P.}\ \bibnamefont {Tsai}},\ }\bibfield  {title} {\enquote {\bibinfo
  {title} {High-efficiency broadband meta-hologram with polarization-controlled
  dual images},}\ }\href@noop {} {\bibfield  {journal} {\bibinfo  {journal}
  {Nano Lett.}\ }\textbf {\bibinfo {volume} {14}},\ \bibinfo {pages} {225}
  (\bibinfo {year} {2014})}\BibitemShut {NoStop}%
\bibitem [{\citenamefont {Pors}\ \emph {et~al.}(2014)\citenamefont {Pors},
  \citenamefont {Nielsen}, \citenamefont {Bernardin}, \citenamefont {Weeber},\
  and\ \citenamefont {Bozhevolnyi}}]{Pors_2014}%
  \BibitemOpen
  \bibfield  {author} {\bibinfo {author} {\bibfnamefont {A.}~\bibnamefont
  {Pors}}, \bibinfo {author} {\bibfnamefont {M.~G.}\ \bibnamefont {Nielsen}},
  \bibinfo {author} {\bibfnamefont {T.}~\bibnamefont {Bernardin}}, \bibinfo
  {author} {\bibfnamefont {J.-C.}\ \bibnamefont {Weeber}}, \ and\ \bibinfo
  {author} {\bibfnamefont {S.~I.}\ \bibnamefont {Bozhevolnyi}},\ }\bibfield
  {title} {\enquote {\bibinfo {title} {Efficient unidirectional
  polarization-controlled excitation of surface plasmon polaritons},}\
  }\href@noop {} {\bibfield  {journal} {\bibinfo  {journal} {Light Sci. Appl.}\
  }\textbf {\bibinfo {volume} {3}},\ \bibinfo {pages} {e197} (\bibinfo {year}
  {2014})}\BibitemShut {NoStop}%
\bibitem [{\citenamefont {Johnson}\ and\ \citenamefont
  {Christy}(1972)}]{johnson}%
  \BibitemOpen
  \bibfield  {author} {\bibinfo {author} {\bibfnamefont {P.~B.}\ \bibnamefont
  {Johnson}}\ and\ \bibinfo {author} {\bibfnamefont {R.~W.}\ \bibnamefont
  {Christy}},\ }\bibfield  {title} {\enquote {\bibinfo {title} {Optical
  constants of the noble metals},}\ }\href@noop {} {\bibfield  {journal}
  {\bibinfo  {journal} {Phys. Rev. B}\ }\textbf {\bibinfo {volume} {6}},\
  \bibinfo {pages} {4370} (\bibinfo {year} {1972})}\BibitemShut {NoStop}%
\bibitem [{\citenamefont {Pancharatnam}(1956)}]{pancharatnam_1956}%
  \BibitemOpen
  \bibfield  {author} {\bibinfo {author} {\bibfnamefont {S.}~\bibnamefont
  {Pancharatnam}},\ }\bibfield  {title} {\enquote {\bibinfo {title}
  {Generalized theory of interference and its applications},}\ }\href@noop {}
  {\bibfield  {journal} {\bibinfo  {journal} {Proc. Indian Acad. Sci. Sect. A}\
  }\textbf {\bibinfo {volume} {44}},\ \bibinfo {pages} {247} (\bibinfo
  {year} {1956})}\BibitemShut {NoStop}%
\bibitem [{\citenamefont {Berry}(1984)}]{berry_1984}%
  \BibitemOpen
  \bibfield  {author} {\bibinfo {author} {\bibfnamefont {M.~V.}\ \bibnamefont
  {Berry}},\ }\bibfield  {title} {\enquote {\bibinfo {title} {Quantal phase
  factors accompanying adiabatic changes},}\ }\href@noop {} {\bibfield
  {journal} {\bibinfo  {journal} {Proc. R. Soc. Lond. A Math. Phys. Sci.}\
  }\textbf {\bibinfo {volume} {392}},\ \bibinfo {pages} {45} (\bibinfo
  {year} {1984})}\BibitemShut {NoStop}%
\bibitem [{\citenamefont {Luo}\ \emph {et~al.}(2015)\citenamefont {Luo},
  \citenamefont {Xiao}, \citenamefont {He}, \citenamefont {Sun},\ and\
  \citenamefont {Zhou}}]{Luo_2015}%
  \BibitemOpen
  \bibfield  {author} {\bibinfo {author} {\bibfnamefont {W.}\ \bibnamefont
  {Luo}}, \bibinfo {author} {\bibfnamefont {S.}\ \bibnamefont {Xiao}},
  \bibinfo {author} {\bibfnamefont {Q.}\ \bibnamefont {He}}, \bibinfo
  {author} {\bibfnamefont {S.}\ \bibnamefont {Sun}}, \ and\ \bibinfo
  {author} {\bibfnamefont {L.}\ \bibnamefont {Zhou}},\ }\bibfield  {title}
  {\enquote {\bibinfo {title} {Photonic spin hall effect with nearly 100\%
  efficiency},}\ }\href@noop {} {\bibfield  {journal} {\bibinfo  {journal}
  {Adv. Opt. Mater.}\ }\textbf {\bibinfo {volume} {3}},\ \bibinfo {pages}
  {1102} (\bibinfo {year} {2015})}\BibitemShut {NoStop}%
\bibitem [{\citenamefont {Foreman}\ and\ \citenamefont
  {T\"or\"ok}(2010)}]{Foreman_2010}%
  \BibitemOpen
  \bibfield  {author} {\bibinfo {author} {\bibfnamefont {M.~R.}\
  \bibnamefont {Foreman}}\ and\ \bibinfo {author} {\bibfnamefont {P.}\
  \bibnamefont {T\"or\"ok}},\ }\bibfield  {title} {\enquote {\bibinfo {title}
  {Information and resolution in electromagnetic optical systems},}\ }\href
  {\doibase 10.1103/PhysRevA.82.043835} {\bibfield  {journal} {\bibinfo
  {journal} {Phys. Rev. A}\ }\textbf {\bibinfo {volume} {82}},\ \bibinfo
  {pages} {043835} (\bibinfo {year} {2010})}\BibitemShut {NoStop}%
\bibitem [{\citenamefont {Twietmeyer}\ and\ \citenamefont
  {Chipman}(2008)}]{Twietmeyer_2008}%
  \BibitemOpen
  \bibfield  {author} {\bibinfo {author} {\bibfnamefont {K.~M.}\ \bibnamefont
  {Twietmeyer}}\ and\ \bibinfo {author} {\bibfnamefont {R.~A.}\ \bibnamefont
  {Chipman}},\ }\bibfield  {title} {\enquote {\bibinfo {title} {Optimization of
  mueller matrix polarimeters in the presence of error sources},}\ }\href
  {\doibase 10.1364/OE.16.011589} {\bibfield  {journal} {\bibinfo  {journal}
  {Opt. Express}\ }\textbf {\bibinfo {volume} {16}},\ \bibinfo {pages}
  {11589} (\bibinfo {year} {2008})}\BibitemShut {NoStop}%
\bibitem [{\citenamefont {Lara}\ and\ \citenamefont
  {Paterson}(2009)}]{Lara_2009}%
  \BibitemOpen
  \bibfield  {author} {\bibinfo {author} {\bibfnamefont {D.}\ \bibnamefont
  {Lara}}\ and\ \bibinfo {author} {\bibfnamefont {C.}\ \bibnamefont
  {Paterson}},\ }\bibfield  {title} {\enquote {\bibinfo {title} {Stokes
  polarimeter optimization in the presence of shot and gaussian noise},}\
  }\href {\doibase 10.1364/OE.17.021240} {\bibfield  {journal} {\bibinfo
  {journal} {Opt. Express}\ }\textbf {\bibinfo {volume} {17}},\ \bibinfo
  {pages} {21240} (\bibinfo {year} {2009})}\BibitemShut {NoStop}%
\bibitem [{\citenamefont {Azzam}\ \emph {et~al.}(1988)\citenamefont {Azzam},
  \citenamefont {Elminyawi},\ and\ \citenamefont {El-Saba}}]{Azzam_1988}%
  \BibitemOpen
  \bibfield  {author} {\bibinfo {author} {\bibfnamefont {R.~M.~A.}\
  \bibnamefont {Azzam}}, \bibinfo {author} {\bibfnamefont {I.~M.}\ \bibnamefont
  {Elminyawi}}, \ and\ \bibinfo {author} {\bibfnamefont {A.~M.}\ \bibnamefont
  {El-Saba}},\ }\bibfield  {title} {\enquote {\bibinfo {title} {General
  analysis and optimization of the four-detector photopolarimeter},}\ }\href
  {\doibase 10.1364/JOSAA.5.000681} {\bibfield  {journal} {\bibinfo  {journal}
  {J. Opt. Soc. Am. A}\ }\textbf {\bibinfo {volume} {5}},\ \bibinfo {pages}
  {681} (\bibinfo {year} {1988})}\BibitemShut {NoStop}%
\bibitem [{\citenamefont {Huang}\ \emph {et~al.}(2013)\citenamefont {Huang},
  \citenamefont {Chen}, \citenamefont {Bai}, \citenamefont {Tan}, \citenamefont
  {Jin}, \citenamefont {Zentgraf},\ and\ \citenamefont {Zhang}}]{Huang_2013}%
  \BibitemOpen
  \bibfield  {author} {\bibinfo {author} {\bibfnamefont {L.}\
  \bibnamefont {Huang}}, \bibinfo {author} {\bibfnamefont {X.}\
  \bibnamefont {Chen}}, \bibinfo {author} {\bibfnamefont {B.}\
  \bibnamefont {Bai}}, \bibinfo {author} {\bibfnamefont {Q.}\
  \bibnamefont {Tan}}, \bibinfo {author} {\bibfnamefont {G.}\ \bibnamefont
  {Jin}}, \bibinfo {author} {\bibfnamefont {T.}\ \bibnamefont {Zentgraf}},
  \ and\ \bibinfo {author} {\bibfnamefont {S.}\ \bibnamefont {Zhang}},\
  }\bibfield  {title} {\enquote {\bibinfo {title} {Helicity dependent
  directional surface plasmon polariton excitation using a metasurface with
  interfacial phase discontinuity},}\ }\href@noop {} {\bibfield  {journal}
  {\bibinfo  {journal} {Light Sci. Appl.}\ }\textbf {\bibinfo {volume} {2}},\
  \bibinfo {pages} {e70} (\bibinfo {year} {2013})}\BibitemShut {NoStop}%
\bibitem [{\citenamefont {Sun}\ \emph {et~al.}(2012)\citenamefont {Sun},
  \citenamefont {He}, \citenamefont {Xiao}, \citenamefont {Xu}, \citenamefont
  {Li},\ and\ \citenamefont {Zhou}}]{Sun_2012}%
  \BibitemOpen
  \bibfield  {author} {\bibinfo {author} {\bibfnamefont {S.}~\bibnamefont
  {Sun}}, \bibinfo {author} {\bibfnamefont {Q.}~\bibnamefont {He}}, \bibinfo
  {author} {\bibfnamefont {S.}~\bibnamefont {Xiao}}, \bibinfo {author}
  {\bibfnamefont {Q.}~\bibnamefont {Xu}}, \bibinfo {author} {\bibfnamefont
  {X.}~\bibnamefont {Li}}, \ and\ \bibinfo {author} {\bibfnamefont
  {L.}~\bibnamefont {Zhou}},\ }\bibfield  {title} {\enquote {\bibinfo {title}
  {Gradient-index meta-surfaces as a bridge linking propagating waves and
  surface waves},}\ }\href@noop {} {\bibfield  {journal} {\bibinfo  {journal}
  {Nat. Mater.}\ }\textbf {\bibinfo {volume} {11}},\ \bibinfo {pages}
  {426} (\bibinfo {year} {2012})}\BibitemShut {NoStop}%
\bibitem [{\citenamefont {Yang}\ \emph {et~al.}(2014)\citenamefont {Yang},
  \citenamefont {Wang}, \citenamefont {Moitra}, \citenamefont {Kravchenko},
  \citenamefont {Briggs},\ and\ \citenamefont {Valentine}}]{Yang_2014}%
  \BibitemOpen
  \bibfield  {author} {\bibinfo {author} {\bibfnamefont {Y.}\ \bibnamefont
  {Yang}}, \bibinfo {author} {\bibfnamefont {W.}\ \bibnamefont {Wang}},
  \bibinfo {author} {\bibfnamefont {P.}\ \bibnamefont {Moitra}},
  \bibinfo {author} {\bibfnamefont {I.~I.}\ \bibnamefont {Kravchenko}},
  \bibinfo {author} {\bibfnamefont {D.~P.}\ \bibnamefont {Briggs}}, \ and\
  \bibinfo {author} {\bibfnamefont {J.}\ \bibnamefont {Valentine}},\
  }\bibfield  {title} {\enquote {\bibinfo {title} {Dielectric meta-reflectarray
  for broadband linear polarization conversion and optical vortex
  generation},}\ }\href@noop {} {\bibfield  {journal} {\bibinfo  {journal}
  {Nano Lett.}\ }\textbf {\bibinfo {volume} {14}},\ \bibinfo {pages}
  {1394} (\bibinfo {year} {2014})}\BibitemShut {NoStop}%
\bibitem [{\citenamefont {Neutens}\ and\ \citenamefont
  {Van~Dorpe}(2013)}]{Neutens_2013}%
  \BibitemOpen
  \bibfield  {author} {\bibinfo {author} {\bibfnamefont {P.}\ \bibnamefont
  {Neutens}}\ and\ \bibinfo {author} {\bibfnamefont {P.}\ \bibnamefont
  {Van~Dorpe}},\ }\enquote {\bibinfo {title} {Integrated plasmonic
  detectors},}\ in\ \href {\doibase 10.1002/9781118634394.ch7} {\emph {\bibinfo
  {booktitle} {Active Plasmonics and Tuneable Plasmonic Metamaterials}}}\
  (\bibinfo  {publisher} {John Wiley \& Sons, Inc.},\ \bibinfo {year} {2013})\
  pp.\ \bibinfo {pages} {219--241}\BibitemShut {NoStop}%
\bibitem [{\citenamefont {Han}\ \emph {et~al.}(2015)\citenamefont {Han},
  \citenamefont {Radko}, \citenamefont {Mazurski}, \citenamefont {Desiatov},
  \citenamefont {Beermann}, \citenamefont {Albrektsen}, \citenamefont {Levy},\
  and\ \citenamefont {Bozhevolnyi}}]{Zhanghua_2015}%
  \BibitemOpen
  \bibfield  {author} {\bibinfo {author} {\bibfnamefont {Z.}\
  \bibnamefont {Han}}, \bibinfo {author} {\bibfnamefont {I.~P.}\ \bibnamefont
  {Radko}}, \bibinfo {author} {\bibfnamefont {N.}\ \bibnamefont {Mazurski}},
  \bibinfo {author} {\bibfnamefont {B.}\ \bibnamefont {Desiatov}}, \bibinfo
  {author} {\bibfnamefont {J.}\ \bibnamefont {Beermann}}, \bibinfo {author}
  {\bibfnamefont {O.}\ \bibnamefont {Albrektsen}}, \bibinfo {author}
  {\bibfnamefont {U.}\ \bibnamefont {Levy}}, \ and\ \bibinfo {author}
  {\bibfnamefont {S.~I.}\ \bibnamefont {Bozhevolnyi}},\ }\bibfield  {title}
  {\enquote {\bibinfo {title} {On-chip detection of radiation guided by
  dielectric-loaded plasmonic waveguides},}\ }\href {\doibase
  10.1021/nl5037885} {\bibfield  {journal} {\bibinfo  {journal} {Nano Lett.}\
  }\textbf {\bibinfo {volume} {15}},\ \bibinfo {pages} {476} (\bibinfo
  {year} {2015})}\BibitemShut {NoStop}%
\end{thebibliography}

%

\end{document}